\documentclass[10pt,reprint,aip,amsmath,amssymb]{revtex4-1}
\usepackage{graphicx}
\input epsf

\newcommand\Alfven{Alfv\'en }
\newcommand\Alfvenic{Alfv\'enic }
\newcommand{\V}[1]{\mathbf{#1}} 
\newcommand{\T}[1]{\texttt{#1}} 
 
\newcommand{\zhat}{\mbox{$\hat{\mathbf{z}}$}}

\newcommand{\secref}[1]{\S\ref{#1}}

\begin{document}

\title{A Prospectus on Kinetic Heliophysics}

 \author{Gregory~G. Howes}
\email[]{gregory-howes@uiowa.edu}
\affiliation{Department of Physics and Astronomy, University of Iowa, Iowa City, 
Iowa 52242, USA.}

\date{\today}

\begin{abstract}
Under the low density and high temperature conditions typical of
heliospheric plasmas, the macroscopic evolution of the heliosphere is
strongly affected by the kinetic plasma physics governing fundamental
microphysical mechanisms.  Kinetic turbulence, collisionless magnetic
reconnection, particle acceleration, and kinetic instabilities are
four poorly understood, grand-challenge problems that lie at the new
frontier of kinetic heliophysics.  The increasing availability of high
cadence and high phase-space resolution measurements of particle
velocity distributions by current and upcoming spacecraft missions and
of massively parallel nonlinear kinetic simulations of weakly
collisional heliospheric plasmas provides the opportunity to transform
our understanding of these kinetic mechanisms through the full
utilization of the information contained in the particle velocity
distributions. Several major considerations for future investigations
of kinetic heliophysics are examined.  Turbulent dissipation followed
by particle heating is highlighted as an inherently two-step process
in weakly collisional plasmas, distinct from the more familiar case in
fluid theory. Concerted efforts must be made to tackle the big-data
challenge of visualizing the high-dimensional (3D-3V) phase space of
kinetic plasma theory through physics-based reductions.  Furthermore,
the development of innovative analysis methods that utilize full
velocity-space measurements, such as the field-particle correlation
technique, will enable us to gain deeper insight into these four
grand-challenge problems of kinetic heliophysics.  A systems approach
to tackle the multi-scale problem of heliophysics through a rigorous
connection between the kinetic physics at microscales and the
self-consistent evolution of the heliosphere at macroscales will
propel the field of kinetic heliophysics into the future.
\end{abstract}


\maketitle 

\section{Introduction}

Humanity continually strives to understand its environment, not only
to ensure its continued survival, but also for the sake of knowledge
itself.  The heliosphere---the realm of influence of our Sun, within
which the planets of our solar system orbit---is our home in the
universe. Nuclear fusion within the core of the Sun is the source of
energy that enables life to thrive on our planet. The majority of this
energy emerges as light, but a small fraction of this energy also
drives a supersonic flow of diffuse ionized gas, or plasma, that blows
radially outward toward the outer reaches of the heliosphere.
Carrying along with it an embedded magnetic field, this solar wind
varies dramatically in response to conditions at the Sun, and is
strongly disturbed during periods of violent activity on the Sun's
surface.  It is the dynamics of this magnetized plasma that governs the
interaction of the Sun with the Earth and the other planets of our
solar system. Humankind has spent billions of dollars to launch
spacecraft to explore our heliosphere in the scientific endeavor to
understand and predict the dynamics of the interplanetary plasma that
affect the Earth and its environment, and here I highlight some issues
at the frontier of that effort.

The inaugural Ronald C. Davidson Award for Plasma Physics recognized
the development of a simple analytical model and supporting numerical
simulations of the turbulent cascade and its kinetic dissipation at
small scales, \citep{Howes:2011b} but to progress further requires
kinetic theory.  The low density and high temperature conditions of
the plasma that fills the heliosphere---as well as many more remote
astrophysical systems---lead to a mean free path for collisions that
is often longer than the length scales relevant to many dynamical
processes of interest. Under these weakly collisional conditions, the
dynamics of the plasma require investigation using the equations of
kinetic plasma physics.

For example, many space and astrophysical plasmas are found to be
turbulent. One of the key impacts of this turbulence is the nonlinear
transfer of the energy of large-scale electromagnetic fields and
plasma flows down to small scales at which the turbulent energy is
ultimately converted to plasma heat, or to some other energization of
the plasma ions and electrons.  At the largest scales of the
astrophysical turbulent cascade in the interstellar
medium,\citep{Armstrong:1981,Armstrong:1995} the length scales of the
turbulent fluctuations may be much larger than the collisional mean
free path, meaning that a fluid description of the turbulent dynamics
at these scales is generally sufficient. But at all scales of the
turbulence in the solar wind,\citep{Bruno:2005,Marsch:2006} and at the
smallest scales of the turbulence in the interstellar
medium,\cite{Spangler:1990} the length scales of the turbulent
fluctuations are much smaller than the collisional mean free path.
Under these conditions, the effect of collisions is negligible on the
timescale of the turbulent fluctuations.  Not only are collisions
insufficient to maintain the Maxwellian particle velocity
distributions that motivate the use of a fluid description, but
collisionless interactions generally dominate the energy exchange
between the fluctuating electromagnetic fields and the plasma
particles.\cite{Howes:2015b,Howes:2017a} Therefore, the equations of
kinetic plasma physics are essential to describe the mechanisms
responsible for removing energy from the turbulent fluctuations and
consequently energizing the plasma particles.

For the weakly collisional interplanetary plasma, such
``microphysical'' kinetic processes govern the heating of the plasma
and the energization of particles, and thereby they exert a
significant influence on the macroscopic evolution of the heliosphere.
Plasma turbulence, magnetic reconnection, particle acceleration, and
instabilities are four fundamental plasma processes operating under
weakly collisional conditions that significantly impact the evolution
of the heliosphere.  These four grand-challenge topics lie at the
frontier of heliophysics. The details of these kinetic plasma
processes remain relatively poorly understood, motivating the
heliophysics community to pursue a coordinated effort of spacecraft
observations, numerical simulations, kinetic plasma theory, and even
laboratory experiments to develop a thorough understanding, and
ultimately a predictive capability, of these processes in kinetic
heliophysics.  This prospectus examines important issues in our
exploration of the kinetic plasma physics of the heliosphere.


\subsection{The Transport of Mass, Momentum, and Energy in the Heliosphere}
The key impact of these four fundamental kinetic plasma physics
processes---kinetic turbulence, collisionless magnetic reconnection,
particle acceleration, and instabilities---is their effect on the
transport of particles, transfer of momentum, and flow of energy
throughout the heliosphere.

Extreme space weather illustrates concisely how the transport of mass,
momentum, and energy by these kinetic plasma physics mechanisms
governs conditions within the heliosphere, possibly leading to adverse
impacts on the Earth and its near-space environment. Magnetic buoyancy
instabilities cause the strong magnetic fields generated by the solar
magnetic dynamo to rise out of the turbulently boiling solar
convection zone, emerging through the photosphere and building up
strong magnetic fields in lower solar atmosphere, or
corona. Eventually, some type of explosive instability can initiate
vigorous magnetic reconnection, hurling tons of magnetized plasma out
into the heliosphere at thousands of kilometers per second, an event
known as a coronal mass ejection. Magnetic energy released through the
process of reconnection can also accelerate electrons back down
towards the photosphere, often causing a powerful solar flare that
enhances x-ray and UV fluxes radiating from the Sun.  In addition, as
the magnetized cloud of ejected plasma barrels at supersonic and
super-\Alfvenic speeds through the slower ambient solar wind, a
collisionless shock forms on the leading edge, frequently accelerating
protons, electrons, and minor ions to nearly the speed of light,
showering the heliosphere in a solar energetic particle event.

These energetic particles stream through the heliosphere, being
scattered by fluctuations in the turbulent interplanetary magnetic
field. Because these energetic particles pose a serious hazard to
communication and navigation satellites as well as manned spacecraft
missions, predicting their fluxes in the near-Earth environment is a
critical element of space weather forecasting, requiring an
understanding of the transport of these particles through the
turbulent solar wind.  In addition, the enhanced x-ray and UV fluxes
from a strong solar flare can boost ionization in the ionosphere,
interfering with or even totally disrupting radio communications with
satellites and aircraft on polar flight paths.

If the coronal mass ejection is directed towards the Earth, its
momentum can lead to a severe compression of the Earth's
magnetosphere, altering the system of currents that modify the Earth's
magnetic field, and triggering a geomagnetic storm. During a
geomagnetic storm, the magnetic field embedded within the ejected
coronal plasma can undergo reconnection with Earth's protective
magnetic field, greatly enhancing the penetration of interplanetary
plasma into the magnetosphere, thereby boosting the density of the
ring current caused by the azimuthal (longitudinal) drift of ions and
electrons trapped in Earth's dipolar magnetic field.  During
particular strong geomagnetic storms, this enhancement of the ring
current can depress the magnitude of the magnetic field at the Earth's
surface by a few percent, causing intense geomagnetically induced
currents that may damage critical components of the electrical power
grid. As the geomagnetic storm rages, the aurorae at the poles light
up, driven either by particles streaming down along open field lines
toward the ionosphere or by the acceleration of electrons by \Alfven
waves which transmit shifts in Earth's distant magnetosphere along
field lines down to the Earth.

This complicated interplay of the different phenomena that constitute
space weather illustrates the fundamental importance of turbulence,
magnetic reconnection, particle acceleration, and instabilities to the
dynamics of the heliosphere and its impact on Earth and society.  It
is important to emphasize that most of the processes mentioned above
remain poorly understood in detail.  An overarching aim of
heliophysics is to improve our understanding of these fundamental
processes and their effect on the transport of particles, momentum,
and energy, with the ultimate aim to develop a predictive capability
for space weather and its impact on our lives.  The path forward is
through the application of kinetic plasma physics to the study of
heliospheric processes, giving birth to the new frontier of
\emph{kinetic heliophysics}.

\subsection{A Coordinated Approach}

Although spacecraft missions enable \emph{in situ} measurements of the
fluctuating electric field $\V{E}$ and magnetic field $\V{B}$ and of
the particle velocity distribution functions in the three-dimensions
of velocity space $f_s(\V{v})$, many of these fundamental kinetic
processes in heliospheric plasmas remain poorly understood. One reason
is that spacecraft measurements suffer the significant limitation that
we measure information only at a single point, or at most a few
points, in space. To circumvent this limitation of spacecraft
observations, many of these kinetic processes can alternatively be
explored in laboratory experiments under more controlled conditions
and with the ability to make measurements at many points in space,
even if it is not possible to achieve the same plasma parameters or
scale separations found in space. A further complication in exploring
kinetic heliophysics is the inherent high dimensionality of kinetic
plasma theory, with its fundamental variables being the particle
distribution functions for each species $s$ in six-dimensional phase
space (3D-3V, three dimensions in physical space and three dimensions
in velocity space), $f_s(\V{r},\V{v},t)$. Theoretical insights from
kinetic plasma theory are vital to reduce this six-dimensional phase
space to a more tractable, smaller number of essential dimensions, for
either space-based or laboratory investigations. Finally, kinetic
numerical simulations provide a critical bridge between the often
idealized conditions susceptible to analytical theory and the more
complex, nonlinear evolution of actual space or laboratory plasmas.

A closely coordinated approach of analytical theory, numerical
simulations, spacecraft measurements, and laboratory experiments has
the greatest potential for transforming our understanding of the
kinetic plasma physics that influences the evolution of the heliosphere.
Here we discuss some important considerations for the next generation
of investigations into kinetic heliophysics.

\section{Damping, Dissipation, and Heating in Weakly Collisional Plasmas}
\label{sec:diss}
A subtle but important issue arises in the investigation of the
conversion of the electromagnetic energy of fields and the kinetic
energy of bulk plasma flows into plasma heat by kinetic physical
mechanisms in weakly collisional heliospheric plasmas. That bottom
line is that, unlike in the more well-known case of fluid systems, in
weakly collisional plasmas the dissipation of turbulent energy into
plasma heat is inherently a two-step process.

Fluid systems are derived from the strongly collisional, or small mean
free path, limit of the Boltzmann equation in kinetic theory.  In this
limit, frequent microscopic collisions maintain the Maxwellian equilibrium
velocity distributions of local thermodynamic equilibrium. A hierarchy
of moment equations may be derived in the limit of small mean free
path (relative to the characteristic length scales of gradients in the
system) by the Chapman-Enskog procedure\cite{Chapman:1970} for neutral
gases, or an analogous procedure for plasma systems.\cite{Grad:1963}
Microscopic collisions in the limit of finite mean free path give rise
to the diffusion of velocity fluctuations by viscosity and of magnetic
field fluctuations by resistivity. Because viscosity and resistivity
are ultimately collisional, the diffusion of the velocity and magnetic
field fluctuations by these mechanisms is irreversible, dissipating
the kinetic and electromagnetic energy of these fluctuations, and
consequently realizing thermodynamic plasma heating and the associated
increase of the system entropy. This picture of plasma heating, based
on the physical intuition derived from the  fluid system, implies that
energy removed from the velocity and electromagnetic field
fluctuations through viscosity and resistivity is immediately
converted into plasma heat.

But in weakly collisional plasmas, the removal of energy from the
electromagnetic field fluctuations and bulk plasma flows is a separate
process from the irreversible conversion of that energy into plasma
heat.  In fact, the energy removed by kinetic processes may not all be
irreversibly converted into heat, but rather some energy may be
channeled instead into nonthermal particle energization, such as the
acceleration of small fraction of particles to high energy, in
apparent defiance of the first law of thermodynamics. These subtleties
require a significantly different approach to the study of the
dissipation of plasma turbulence and the resulting energization of the
plasma under the typically weakly collisional conditions of
heliospheric plasmas.

To consider in more detail the dynamics and dissipation of turbulence
in weakly collisional heliospheric plasmas, we turn to the Boltzmann
equation which governs the evolution of the six-dimensional velocity
distribution function $f_s(\V{r},\V{v},t)$ for a plasma species $s$,
\begin{equation}
\frac{\partial f_s}{\partial t} + \mathbf{v}\cdot \nabla f_s + \frac
     {q_s}{m_s}\left[ \mathbf{E}+ \frac{\mathbf{v} \times \mathbf{B}
       }{c} \right] \cdot \frac{\partial f_s}{\partial \mathbf{v}} =
     \left(\frac{\partial f_s}{\partial t} \right)_{\mbox{coll}}.
\label{eq:boltzmann}
\end{equation}
Combining a Boltzmann equation for each species with Maxwell's
equations forms the closed set of Maxwell-Boltzmann equations that
govern the nonlinear kinetic evolution of a plasma.  In the inner
heliosphere (within 1~AU of the sun), the typical conditions of the
interplanetary plasma lead to a collisional mean free path that is of
order 1~AU, approximately $10^8$~km.\cite{Marsch:2006} In comparison,
the largest scale structures of the interplanetary turbulent cascade
have a length scale of  $10^6$~km. The upshot is that the collisional term
on the right-hand side of \eqref{eq:boltzmann} is subdominant, not
significantly affecting the turbulent dynamics on the timescale of the
turbulent fluctuations.

Since the collisional term in \eqref{eq:boltzmann} is insufficient to
diminish the turbulent fluctuations in heliospheric plasmas, the
removal of energy from the turbulent electromagnetic field and bulk
plasma flow fluctuations occurs through interactions between the
electromagnetic fields and the charged plasma particles,
\cite{Klein:2016a,Howes:2017a} and these interactions are governed by
the Lorentz force term, the third term on the left-hand side of
\eqref{eq:boltzmann}. The linear collisionless wave-particle
interactions---such as Landau damping,\cite{Landau:1946,Villani:2014}
Barnes damping,\cite{Barnes:1966} and cyclotron
damping\cite{Stix:1992}---provide familiar examples of such
interactions.  But it is important to note that a net transfer of
energy from fields to particles, depleting the energy of
electromagnetic fluctuations and boosting the microscopic kinetic
energy of the particles, can occur under more general circumstances
that do not require the persistent presence of waves. Fundamentally,
when collisions are weak, the only avenue to remove energy from
electromagnetic field and bulk plasma flow fluctuations is through
collisionless interactions between the fields and the particles, where
the electromagnetic forces do net work on the plasma particles.

One of the key fundamental distinctions, compared to the viscous and
resistive dissipation in a fluid system, is that the net energy
transfer between fields and particles by the work of electromagnetic
forces is reversible, with no associated increase in the system
entropy. In a kinetic system, Boltzmann's $H$ Theorem shows that the
increase of entropy, and therefore irreversible plasma heating, can
only be accomplished through
collisions.\cite{Howes:2006,Schekochihin:2009} So, how can one achieve
irreversible heating in a plasma of arbitrarily weak collisionality?
To accomplish irreversible heating requires a subsequent process that
enhances the effectiveness of even arbitrarily weak collisions, as
explained below.

When the collisionless interaction between fields and particles
mediates the removal of energy of the electromagnetic field
fluctuations, that energy is transferred to the particles, appearing
as fluctuations in velocity space of the particle distribution
functions.  The general form of the collision operator involves
second-derivatives in velocity, so the rate of change of the
distribution function due to collisions takes the
form\citep{Schekochihin:2009} $\partial f /\partial t \sim \nu \ v_t^2
\partial^2 f /\partial v^2 \sim \nu/(\Delta v/v_t)^2 f$. We must
compare the rate of the collisional evolution to the typical frequency
$\omega$ of the turbulent fluctuations, governed by the other terms of
the Boltzmann equation. Even if the collisional frequency $\nu$ is
arbitrarily small, $\nu \ll \omega$, the rate of change of the
distribution function due to collisions can compete with the frequency
of turbulent fluctuations if the scale of the velocity space
fluctuations $\Delta v$ is sufficiently small relative to the typical
thermal velocity $v_t$, $\Delta v/v_{t} \sim (\nu/\omega)^{1/2}$.
Note that these small fluctuations in velocity space typically
contribute little to the first moment of the distribution functions
(which yields the bulk plasma flows and current density), so the
turbulent fluctuations are insignificantly affected by these small
fluctuations in velocity space.  \cite{Howes:2015b}

How do the fluctuations generated by collisionless interactions reach
sufficiently small scales in velocity space that collisions can
effectively smooth them out?  The answer depends on the associated
spatial length scale of the fluctuations. One process is the linear
phase mixing governed by the ballistic term of the Boltzmann equation
(the second term on the left-hand side of \eqref{eq:boltzmann}). Note,
however, it has been recently suggested that, for length scales large
relative to the thermal Larmor radius of particle species $s$,
$k_\perp \rho_s \ll 1$, an anti-phase mixing mechanism in the presence
of turbulence may prevent these fluctuations in velocity space from
reaching sufficiently small velocity scales, $\Delta v/v_{ts} \sim
(\nu/\omega)^{1/2}$, to be thermalized by weak collisions.
\cite{Schekochihin:2016,Parker:2016} At length scales smaller than the
Larmor radius, $k_\perp \rho_s \gtrsim 1$, a nonlinear phase-mixing
mechanism, arising from differential drifts due to the
particle-velocity-dependent Larmor averaging of the electromagnetic
fields,\cite{Dorland:1993} also known as the entropy cascade,
\cite{Schekochihin:2009,Tatsuno:2009,Plunk:2010,Plunk:2011,Kawamori:2013}
may effectively drive velocity-space fluctuations to sufficiently
small scales to achieve irreversible heating through collisions.

The primary message here is that the physical mechanisms governing the
damping of turbulent fluctuations and the subsequent irreversible
heating in weakly collisional heliospheric plasmas has an inherently
different nature from dissipation in the strongly collisional, fluid
systems with which most people are more familiar.  Kinetic plasma
physics plays a central role in the process of particle energization,
defining a key frontier in kinetic heliophysics.  Below, we will
highlight how these key differences motivate powerful new approaches
to the study of the flow of energy throughout the heliosphere,
approaches that fully utilize the measurements routinely made by
modern spacecraft missions.

\section{Velocity Space: The Next Frontier}
Tackling the six-dimensional phase space of kinetic plasma physics
presents the new challenge of interpreting not only the fluctuations
in space and time, as necessary in fluid theory as well, but also the
dynamics in velocity space. By utilizing the full information content
of velocity-space measurements, however, we have the tremendous
opportunity to realize a transformative leap in our understanding of
kinetic heliophysics. Visualization of the high-dimensional datasets
of modern spacecraft instrumentation and cutting-edge kinetic
numerical simulations represents a new, big-data challenge.
Physics-driven reduction of the data is essential for interpreting the
results of complicated nonlinear kinetic dynamics, and innovative new
analysis methods promise to shed new light on how particles in
different regions of velocity space contribute to the dynamics.  Here
I present some thoughts on exploiting velocity space, the next
frontier in kinetic heliophysics.

\subsection{New Insights Lurking in Velocity Space}
Spacecraft suffer the inherent limitation that measurements are made
at only a single point in space (or in the case of multi-spacecraft
missions, a few points in space). But, at that single point in space,
ion and electron instruments can measure the full three-dimensional
distribution of particle velocities. Velocity space is a messy place,
especially in the turbulent state typical of heliospheric plasmas, and
although the fluctuations in the particle velocity distribution
functions are hard to interpret, they contain a vast store of
information that has been largely underutilized.

Often spacecraft measurements of the velocity distributions are used
to compute moments of the distributions, yielding the density, bulk
flow velocity, and (possibly anisotropic) temperature of the plasma,
while more sophisticated approaches may compute other dynamic
quantities, such as the heat flux. For example, over the last fifteen
years, several breakthrough observational investigations have
illuminated the role of kinetic temperature anisotropy instabilities
in regulating the temperature anisotropy of the solar wind plasma.
\citep{Kasper:2002,Hellinger:2006,Matteini:2007,Bale:2009} Kinetic
plasma theory \citep{Gary:1976} provided critical guidance in this
case, suggesting that the action of the these kinetic temperature
anisotropy instabilities is most clearly illustrated on a plot of the
$(\beta_\parallel, T_\perp/T_\parallel)$ plane, often called a Brazil
plot because the distribution of measurements of the near-Earth solar
wind plasma on this plane resembles the geographic outline of Brazil.

Velocity space, however, contains far more information about the
kinetic dynamics of heliospheric plasmas than just these low-order
moments. In particular, velocity space retains an imprint of the
collisionless interactions between the electromagnetic fields and the
plasma particles, so the investigation of the morphology of the
velocity distribution functions can be used to gain insight into the
processes which govern the plasma evolution.

Early measurements from the \emph{Helios} spacecraft within 1~AU
showed proton velocity distributions with a strongly anisotropic core
(having a characteristic temperature perpendicular to the local
magnetic field that is greater than the temperature parallel to the
field) and a significant field-aligned beam. \cite{Marsch:1982}
Subsequent detailed examinations of the equilibrium proton velocity
distribution functions measured in the solar wind have sought evidence
for the quasilinear diffusion of proton distribution functions through
pitch angle scattering by ion cyclotron waves
\citep{Marsch:2001,Tu:2002,Heuer:2007,Marsch:2011,He:2015a} and for
the development of a plateau (quasilinear flattening) in the
distribution function along the field-aligned direction through Landau
damping.  \citep{He:2015a} Currently, proton and electron velocity
distribution functions measured at unprecedented phase-space
resolution and cadence by the \emph{Magnetospheric Multiscale}
(\emph{MMS}) mission \citep{Burch:2016a} are providing a detailed view
of the kinetic plasma dynamics associated with collisionless magnetic
reconnection in the Earth's magnetosphere.\citep{Burch:2016}

Indeed, searching for evidence of the quasilinear evolution of the
mean velocity distribution functions by examining structures in
velocity space can provide important clues about the kinetic evolution
of the plasma, but there is actually much more information contained
within the \emph{fluctuations} in velocity space.

For example, the Morrison $G$ transform \citep{Morrison:1994} is an
integral transform of the perturbations in the velocity distribution
function for an electrostatic system. This transform enables the
perturbation to the distribution function to be written as a weighted
sum of Case-Van Kampen modes, a continuous spectrum of solutions to
the Vlasov equation. \citep{VanKampen:1955,Case:1959} With reasonable
assumptions, the Morrison $G$ transform can be exploited to
reconstruct the spatial dependence of the electric field from
measurements of the perturbed distribution function made at just a
single location in space.\citep{Skiff:2002} This example illustrates
the potential for fully exploiting the information contained in the
fluctuations in velocity space to gain much deeper insight into the
kinetic plasma dynamics, an approach that requires detailed guidance
from kinetic plasma theory.

Transformative progress can be made in kinetic heliophysics by
capitalizing on the power of kinetic plasma theory to devise
insightful new analysis techniques that can be applied to the high
cadence and high phase-space resolution measurements of particle
velocity distributions enabled by modern spacecraft instrumentation.
One such promising new method is the field-particle correlation
technique, \citep{Klein:2016a,Howes:2017a} described below in
\secref{sec:fpcorr}.  Cutting-edge nonlinear kinetic simulations of
the plasma dynamics provide a valuable tool both to test these new
techniques under realistic plasma conditions and to interpret the
results of their application to spacecraft measurements.  Finally, the
development of powerful new diagnostics for measuring the velocity
distribution functions in the laboratory will enable complementary
experiments that test critical aspects of kinetic physical processes
in space plasmas.

\subsection{Visualizing Velocity Space}

A key challenge for fully utilizing velocity distribution measurements
is the visualization and analysis of the high-dimensional data arising
from the six-dimensional (3D-3V) phase space of kinetic plasma theory.
In particular, nonlinear kinetic numerical simulations are able to
compute the full six-dimensional velocity distribution functions for
each species, $f_s(\V{r},\V{v},t)$, at each point in time, resulting
in a big-data challenge for the analysis of kinetic heliophysics
problems. Six-dimensions is more than can easily visualized, so a
physics-driven reduction of this high-dimensional data is essential
for the interpretation of the complicated nonlinear kinetic dynamics.

Even for spacecraft observations, where particle velocities are
measured at only a single point in space as a function of time,
visualizing the three-dimensional velocity distributions can be
awkward, but theoretical considerations can point to helpful
simplifications.  The recent study by He \emph{et al.},
\citep{He:2015a} for example, presents cross-sections through the
three-dimensional proton velocity distribution functions measured by
the \emph{WIND} spacecraft. Physical considerations led them to orient
these cross-sections relative to the directions of the solar wind flow
velocity and the local magnetic field, enabling the characteristic
structures in the mean velocity distribution functions to be more
easily seen.

But rather than taking cross-sections through a three-dimensional
velocity space, which effectively discards the bulk of the 3-V
information that lies outside of that cross-section, integrating the
data over an ignorable coordinate incorporates the full data set,
yielding an improved signal-to-noise ratio.  Under the strongly
magnetized conditions typical of heliospheric plasmas---specifically
meaning that the typical radius of a particle's Larmor motion about the
magnetic field is much smaller than the length scale of spatial
gradients in the plasma equilibrium, a condition that is almost always
well satisfied in space plasmas\citep{Howes:2006}---the local magnetic
field establishes a preferred direction in the plasma.  In this case,
the helical motion of a charged particle about the magnetic field,
caused by the Lorentz force, is most efficiently expressed using
cylindrical coordinates for velocity space,
$(v_\perp,\theta,v_\parallel)$. Here $v_\perp$ is the velocity
perpendicular to the local magnetic field, $\theta$ is the angle of
the particle's gyromotion about the magnetic field, and $v_\parallel$
is the particle velocity parallel to the local magnetic field.

If the characteristic frequencies for the evolution of both the
equilibrium and the fluctuations are smaller than the cyclotron
frequency, $\omega \ll \Omega$, then the distribution function turns
out to be \emph{gyrotropic}, \citep{Schekochihin:2010} meaning that it
is independent of the gyrophase angle $\theta$ about the magnetic
field, $f(v_\perp, v_\parallel)$.  Therefore, integrating over the
gyrophase angle incorporates all of the data in three-dimensional
velocity space to yield an optimal representation in \emph{gyrotropic
  velocity space}, $(v_\perp, v_\parallel)$. Note that, in the case of
spacecraft measurements, the origin of the velocity-space coordinate
system should be centered at the plasma bulk flow velocity.

It is worthwhile noting that, even in cases where the frequencies of
the fluctuations violate the low frequency approximation, $\omega
\gtrsim \Omega$, and thereby the physics cannot be described using a
gyrotropic model, gyrotropic velocity space $(v_\perp, v_\parallel)$
may still be a useful reduction of the three-dimensional velocity
space for visualization.  For example, in the case of cyclotron
damping, the dynamics are inherently not gyrotropic, but the effect on
the distribution function is a broadening of the distribution in the
plane perpendicular to the local magnetic
field,\citep{Isenberg:2001,Isenberg:2012} an impact that can be
usefully visualized in gyrotropic velocity space $(v_\perp,
v_\parallel)$.

In summary, determining the optimal, physics-based reductions of the
six-dimensional (3D-3V) phase space of kinetic plasma theory for
important heliophysics problems will enable better utilization of the
full information content of velocity space. Tackling this challenge to
visualize efficiently the high-dimensional data will enable the
heliophysics community not only to maximize the scientific return from
the high phase-space resolution plasma measurements of current and
upcoming spacecraft missions, but also to gain deeper insight into the
underlying kinetic physical mechanisms governing the evolution of
massively parallel, nonlinear kinetic numerical simulations.

\subsection{Spectral Decomposition of Velocity Space}
\label{sec:velspectral}

In a weakly collisional plasma, valuable insight into the flux of free
energy in velocity space can be gained by using an appropriate
spectral decomposition of the structure of the perturbations to the
velocity distribution function. The kinetic equation for the evolution
of fluctuations in velocity space parallel to the magnetic field is
simplified by recasting the perturbed distribution functions in terms
of Hermite polynomials, an approach exploited in early investigations
of kinetic plasma physics.
\citep{Armstrong:1967,Grant:1967,Hammett:1993,Parker:1995,Ng:1999,Watanabe;2004}
Specifically, the linear parallel phase mixing due to the ballistic
term in the kinetic equation reduces to a coupling between adjacent
Hermite moments, and the Lenard-Bernstein collision operator
\citep{Lenard:1958} takes on a particularly simple form since its
eigenfunctions are the Hermite polynomials. Recent studies of the
dissipation of weakly collisional plasma turbulence have exploited
this Hermite representation of parallel velocity fluctuations to
diagnose the flow of energy through velocity space.
\citep{Zocco:2011,Hatch:2013b,Loureiro:2013,Hatch:2014,Plunk:2014,Loureiro:2016,Parker:2016,Schekochihin:2016}
Likewise, the perpendicular velocity-space structure arising from
nonlinear phase mixing
\citep{Dorland:1993,Schekochihin:2009,Tatsuno:2009,Plunk:2010} can be
conveniently represented using a Hankel transform,
\citep{Tatsuno:2009,Plunk:2010,Tatsuno:2010,Plunk:2011,Tatsuno:2012}
enabling the flow of energy to smaller scales in perpendicular
velocity to be diagnosed clearly.  Further use of these optimal
spectral decompositions of the structure of fluctuations in velocity
space will facilitate greater insights into the nature of particle
energization in weakly collisional heliospheric plasmas.

\subsection{Field-Particle Correlations}
\label{sec:fpcorr}
To make the most of the velocity-space information provided by modern
spacecraft instrumentation and high-performance kinetic numerical
simulations, it is essential to develop innovative analysis methods
that enable us to gain deeper insight into the grand-challenge
problems of kinetic heliophysics: turbulence, collisionless magnetic
reconnection, particle acceleration, and instabilities. The recently
developed \emph{field-particle correlation technique}
\citep{Klein:2016a,Howes:2017a} employs the electromagnetic field
fluctuations along with fluctuations in the particle velocity
distribution functions to determine the energy transfer between the
fields and particles.

The idea of using correlated field and particle measurements to
explore the kinetic physics of space plasmas has found limited
application in the aurora
\citep{Ergun:1991a,Ergun:1991b,Muschietti:1994,Ergun:1998,Ergun:2001,Kletzing:2005,Kletzing:2006}
and the Earth's magnetosphere \citep{Gough:1981,Watkins:1996} using
wave-particle correlator instruments flown on sounding rockets and
spacecraft. A detailed review of these previous efforts is found in
Howes, Klein and Li. \citep{Howes:2017a} These early instrumental
efforts largely focused on seeking electron phase-space bunching in
finite amplitude Langmuir waves, in a regime where the electron count
rate was generally significantly lower than the frequency of the
Langmuir waves.

With the modern instrumentation on current
(\emph{MMS}\citep{Burch:2016}), upcoming (\emph{Solar Probe
  Plus}\citep{Bale:2016,Kasper:2016} and \emph{Solar
  Orbiter}\citep{Muller:2013}), and proposed (\emph{Turbulence Heating
  ObserveR}, \emph{THOR}\citep{Vaivads:2016}) spacecraft missions,
particle velocity distribution function measurements can now be made
with unprecedented phase-space resolution and at cadences sufficient
to resolve the frequencies of the electromagnetic turbulent
fluctuations involved in the damping of the turbulence and resulting
energization of the particles.  With access now to such high
quality velocity-space data from spacecraft observations, and with
cutting-edge numerical simulations now capable of simulating the full
high-dimensional phase-space of kinetic plasma physics, advanced
analysis methods based on kinetic plasma theory have the potential to
break new ground in our understanding of kinetic heliophysics.

The field-particle correlation technique was developed to exploit
these new instrumental and computational capabilities to provide a new
window on the kinetic mechanisms at play in heliospheric plasmas.  The
novel aspect of this method is that it determines the energy transfer
between fields and particles as a function of the particle velocity,
yielding a velocity-space signature that characterizes the kinetic
mechanism responsible for the energy transfer.

This technique was primarily developed to diagnose the particle
energization in plasma turbulence as energy is removed from the
turbulent magnetic field and plasma flow fluctuations through
collisionless interactions between the fields and particles.  The
method, however, is simply based on the equations of nonlinear kinetic
plasma theory.\citep{Howes:2017a} At the most basic level,
collisionless magnetic reconnection, particle acceleration, and
kinetic instabilities are simply nonlinear kinetic plasma physics
phenomena, mediated by interactions between the electromagnetic fields
and particles. Therefore, the field-particle correlation approach is a
fundamental way to explore the evolution of these other processes and
their impact on the plasma environment (often significantly
influencing the large-scale, macroscopic evolution of the system).

As emphasized earlier in \secref{sec:diss}, under the weakly
collisional conditions relevant to most heliospheric plasmas, the
collisional term in the Boltzmann equation \eqref{eq:boltzmann} cannot
be responsible for the damping of the turbulent fluctuations. Instead,
the Lorentz force term, the third term on the left-hand side of
\eqref{eq:boltzmann}, governs the collisionless interactions that lead
to the net transfer of energy from the electromagnetic fields to the
microscopic kinetic energy of individual plasma particles. Therefore,
we may drop the collisional term on the right-hand side of
\eqref{eq:boltzmann} to obtain the Vlasov equation for the following
analysis.

As an example of the application of the field-particle correlation
technique, we briefly derive here the appropriate field-particle
correlation for Landau damping in a 3D, electromagnetic plasma.  We
begin by multiplying the Vlasov equation by $m_s v^2/2$ to obtain an
expression for the rate of change of the phase-space energy density,
\begin{eqnarray}
  \lefteqn{ \frac{\partial w_s(\V{r},\V{v},t)}{\partial t} =}& &  \label{eq:dws}\\
&&   - \V{v}\cdot \nabla  w_s  -
  q_s\frac{v^2}{2}  \mathbf{E} \cdot \frac{\partial f_s}{\partial \mathbf{v}}
  - \frac{q_s}{c}\frac{v^2}{2} \left(\mathbf{v} \times \mathbf{B}\right)
      \cdot \frac{\partial f_s}{\partial \mathbf{v}}, \nonumber 
\end{eqnarray}
where the energy density in six-dimensional phase space for a particle
species $s$ is given by $w_s(\V{r},\V{v},t) = m_s v^2 f_s(\V{r},\V{v},t)/2$.

Under appropriate boundary conditions, such as periodic or infinite
spatial boundaries, the the first and third terms on the right-hand side of
\eqref{eq:dws} yield zero net energy transfer upon integration over
all phase-space, including both spatial volume and velocity space.
Therefore, this fundamental application of nonlinear kinetic plasma
theory shows that it is the second term that is responsible for the
net energy transfer between fields and particles in a collisionless plasma.
Since Landau damping is mediated by the component of the electric
field parallel to the local magnetic field, $E_\parallel$, the term
that is responsible for the energy transfer from fields to particles
through Landau damping has the form \citep{Klein:2017b}
\begin{equation}
 -  q_s\frac{v_\parallel^2}{2}\frac{\partial f_s}{\partial v_\parallel}  E_\parallel .
  \label{eq:epar}
\end{equation}
Note that the $v^2=v_\parallel^2+v_\perp^2$ factor is reduced to
$v_\parallel^2$ here because the net energy change is zero for the
$v_\perp^2$ contribution when integrated over velocity.

But the term in \eqref{eq:epar} not only governs the physics of the
net transfer of energy to the particles through the collisionless
Landau damping of the electromagnetic fluctuations, but also contains
a significant contribution from the undamped oscillatory motion in the
plasma that yields no net energization of
particles. \citep{Howes:2017a} To eliminate this contribution of the
oscillatory energy transfer, which often has a larger amplitude than
the secular transfer of energy that does yield a net energization of
particles, we perform a correlation of the two factors in
\eqref{eq:epar} over a suitably chosen correlation interval $\tau$,
\begin{equation}
  C_{E_\parallel} (\V{v},t,\tau)= C\left(- q_s\frac{v_\parallel^2}{2}
  \frac{\partial f_s(\V{r}_0,\V{v},t)}{\partial
    v_\parallel},E_\parallel(\V{r}_0,t)\right)
   \label{eq:cepar}
\end{equation}
This unnormalized correlation gives the phase-space energy transfer
rate between species $s$ and the parallel electric field, and retains
its functional dependence on velocity space.

We emphasize here that this method requires measurements of
$f_s(\V{v},t)$ and $E_\parallel(t)$ at only a single-point in space
$\V{r}_0$.  In order to achieve the cancellation of the oscillatory
energy transfer, the measurements simply need to span at least $2 \pi$
in the phase of the fluctuations.\citep{Howes:2017a} Essentially, this
method is complementary to the approach used in quasilinear theory,
where a spatial integration over all volume is used to eliminate any
oscillatory contribution; here, we integrate over time, rather than
space, to sample the full $2 \pi$ phase of the fluctuations.  Note
however, that in the presence of fluctuations with different
characteristic frequencies (for example, with dispersive waves that
are common in plasma physics, such as kinetic \Alfven waves, or in a
plasma exhibiting broadband turbulent fluctuations), the integration
over time achieves only an approximate cancellation of the oscillatory
component, rather than the exact cancellation that is achieved in
quasilinear theory using integration over all space.

It can be shown, through the integration of term \eqref{eq:epar} over
velocity-space, that the net energy transfer rate to a species $s$ is
equivalent to $j_{\parallel s} E_\parallel$, the rate of net work done
on the particles by the parallel electric field, \citep{Howes:2017a}
an approach previously used with spacecraft observations as a direct
measure of the plasma heating.
\citep{Retino:2007,Sundkvist:2007,Burch:2016} However, by not
integrating over velocity space, the field-particle correlation
technique provides much more information than just the net rate of
energy transfer to the particles---it provides the distribution of
that energy transfer in velocity-space, denoted here the
\emph{velocity-space signature}, potentially enabling different
mechanisms of energy transfer to be distinguished.

\subsection{Example: Velocity-Space Signature of the Landau Damping of a Kinetic \Alfven Wave}
Here I present the application of the field-particle correlation
technique to determine the velocity-space signature of the particle
energization due to the Landau damping of a kinetic \Alfven wave.

A useful reduction of the six-dimensional phase-space of kinetic
plasma theory for the modeling of the Landau damping of kinetic
\Alfven waves is the gyrokinetic approximation. The derivation of
gyrokinetics, a rigorous low-frequency anisotropic limit of kinetic
plasma theory,
\cite{Rutherford:1968,Taylor:1968,Antonsen:1980,Catto:1981,
  Frieman:1982,Dubin:1983,Hahm:1988,Howes:2006,Schekochihin:2009}
systematically averages out the particle cyclotron motion, leading to
a reduction of the three-dimensional velocity space $(v_\perp, \theta,
v_\parallel)$ to the two-dimensional gyrotropic velocity space
$(v_\perp,v_\parallel)$. This procedure orders out the fast
magnetosonic and whistler waves as well as the cyclotron resonances,
but retains finite Larmor radius effects and the collisionless Landau
resonance. We employ here the Astrophysical Gyrokinetics code
\T{AstroGK} \citep{Numata:2010} to perform a nonlinear gyrokinetic
simulation of the Landau damping of a single kinetic \Alfven wave.

\T{AstroGK} evolves the perturbed gyroaveraged distribution function
$h_s(x,y,z,\lambda,\varepsilon)$ for each species $s$, the scalar
potential $\varphi$, the parallel vector potential $A_\parallel$, and the
parallel magnetic field perturbation $\delta B_\parallel$ according to
the gyrokinetic equation and the gyroaveraged Maxwell's
equations.\citep{Frieman:1982,Howes:2006} Velocity space coordinates are
$\lambda=v_\perp^2/v^2$ and $\varepsilon=v^2/2$. The domain is a
periodic box of size $L_{\perp }^2 \times L_{\parallel }$, elongated
along the straight, uniform mean magnetic field $\V{B}_0=B_0 \zhat$,
where all quantities may be rescaled to any parallel dimension
satisfying $L_{\parallel } /L_{\perp } \gg 1$. Uniform Maxwellian
equilibria for ions (protons) and electrons are chosen, with the
correct mass ratio $m_i/m_e=1836$. Spatial dimensions $(x,y)$
perpendicular to the mean field are treated pseudospectrally; an
upwind finite-difference scheme is used in the parallel direction,
$z$. Collisions employ a fully conservative,
linearized collision operator with energy diffusion and
pitch-angle scattering. \citep{Abel:2008,Barnes:2009}

We initialize a single kinetic \Alfven wave with $k_\perp \rho_i=1.3$
for plasma parameters $\beta_i=1$ and $T_i/T_e=1$ in a simulation
domain of size $L_{\perp}=2\pi \rho_i/1.3$ and $L_{\parallel } =
L_{\perp }/\epsilon$, where $\epsilon \ll 1$ is the gyrokinetic
expansion parameter.  The simulation resolution is
$(n_x,n_y,n_z,n_\lambda,n_\varepsilon,n_s)= (10,10,32,64,64,2)$.  The
initialization procedure \citep{Nielson:2013a} specifies the initial
perturbed distribution functions and electromagnetic fields according
to the eigenfunction from the linear collisionless gyrokinetic
dispersion relation \citep{Howes:2006} for the kinetic \Alfven wave.  The solution for
this kinetic \Alfven wave has a linear frequency $\omega/\omega_A =
1.237$ and collisionless damping rate $\gamma /\omega_A = -0.0445$,
yielding a normalized period $T \omega_A = 5.079$, where $\omega_A=
k_\parallel v_A = 2 \pi v_A/L_\parallel$ is the characteristic angular
frequency associated with crossing the parallel domain length
$L_\parallel$ at the the \Alfven speed $v_A$.  The initialization
procedure includes a short linear evolution of five wave periods with
enhanced collisionality $\nu_i=\nu_e=0.02\omega_A$ to eliminate any
transients in the initial conditions that do not satisfy the properties
of the desired kinetic \Alfven wave. After the linear transient
elimination, the nonlinear evolution of the simulation begins with
$\nu_i=\nu_e=0.002\omega_A$, leading to weakly collisional conditions
with $\nu_s/\omega \sim 10^{-3}$.

\begin{figure}
\resizebox{3.25in}{!}{\includegraphics*[1.15in,1.1in][5.1in,3.75in]{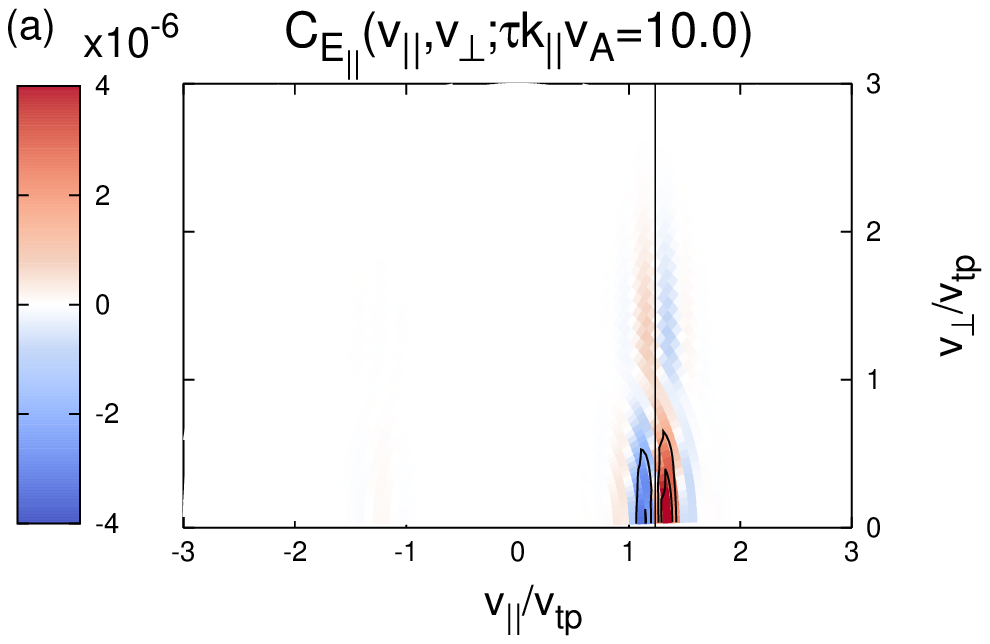}}
\resizebox{3.25in}{!}{\includegraphics*[1.15in,1.1in][5.1in,3.75in]{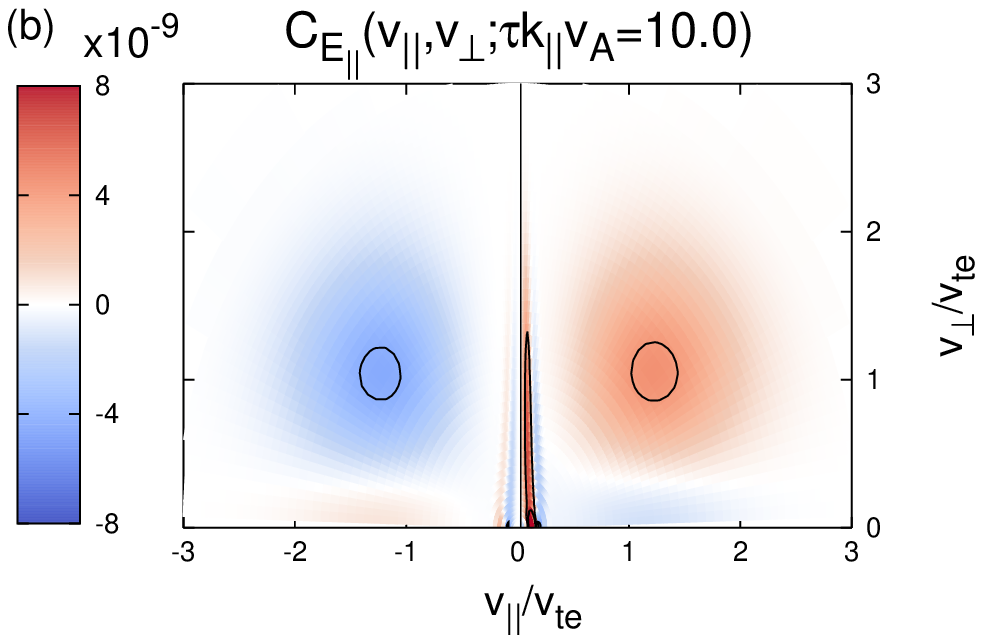}}
\resizebox{3.25in}{!}{\includegraphics*[1.15in,1.1in][5.1in,3.75in]{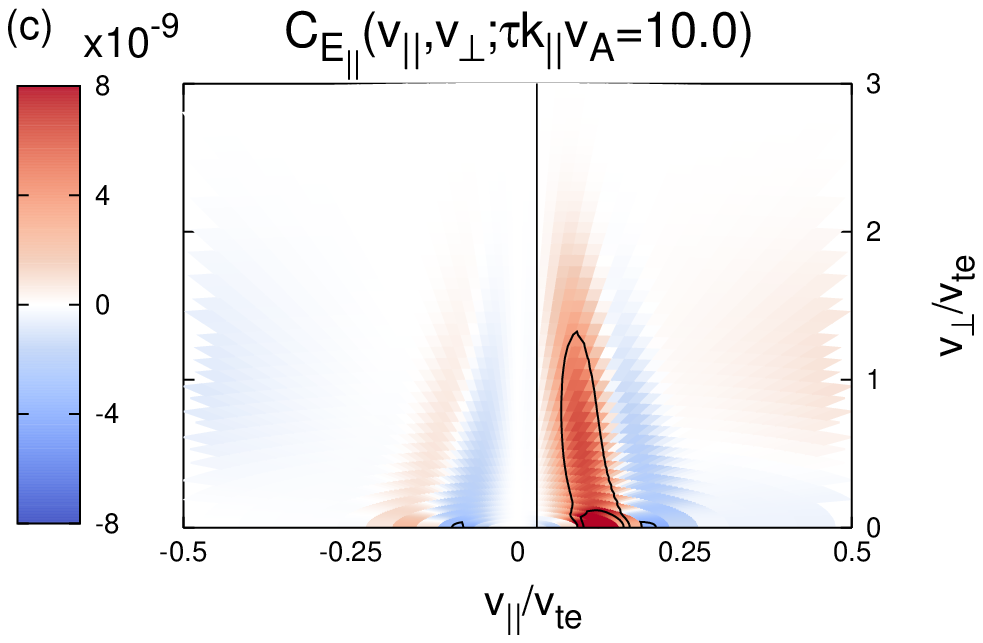}}
\caption{The gyrotropic ($v_\parallel,v_\perp)$ velocity-space
  signature of Landau damping of a kinetic \Alfven wave with $k_\perp
  \rho_i=1.3$, $\beta_i=1$, and $T_i/T_e=1$ using a correlation
  interval $\tau k_\parallel v_A = 10$. (a) The correlation
  $C_{E_\parallel} \left ( - (q_i v_\parallel^2/2) \partial
    f_i(\V{v})/\partial v_\parallel,E_\parallel\right)$ for ions
  shows a clear signature at the resonant velocity,
  $v_\parallel/v_{ti}= \omega/(k_\parallel v_{ti})= 1.237$ (vertical black line). (b) The correlation
  $C_{E_\parallel} \left ( - (q_e v_\parallel^2/2) \partial
    f_e(\V{v})/\partial v_\parallel,E_\parallel\right)$ for
  electrons shows a signature at the resonant velocity
  $v_\parallel/v_{te}= \omega/(k_\parallel v_{te})= 0.029$(vertical black line).
  (c) Zooming into the region $-0.5 \le v_\parallel/v_{te} \le 0.5$
  for $C_{E_\parallel}$ for the electrons, detailing the distribution
  of the energy transfer near the resonant velocity
  $v_\parallel/v_{te}= \omega/(k_\parallel v_{te})= 0.029$.
\label{fig:ld_kaw}}
\end{figure}

As the nonlinear simulation evolves, the distribution functions for
each species and the electromagnetic fields are sampled at one spatial
point in the simulation domain. We choose a correlation interval $\tau
\omega_A = 10.0$, which is approximately equal to two periods of the
kinetic \Alfven wave. In Figure~\ref{fig:ld_kaw}(a), we present the
gyrotropic $(v_\parallel, v_\perp)$ velocity-space signature
$C_{E_\parallel} \left ( - (q_i v_\parallel^2/2) \partial
f_i(\V{v})/\partial v_\parallel,E_\parallel\right)$ for the ions, showing
the localization in velocity space of the energy transfer between the
parallel electric field and the ions around the phase velocity of the
kinetic \Alfven wave, $v_\parallel/v_{ti}= \omega/(k_\parallel
v_{ti})= 1.237$ (vertical black line). Similar to the case of the
Landau damping of electrostatic fluctuations (Langmuir waves) in a
1D-1V Vlasov-Poisson plasma,\citep{Klein:2016a,Howes:2017a} the energy
gain (red) by ions with $v> \omega/k_\parallel$ and energy loss (blue)
by ions with $v< \omega/k_\parallel$ is a signature of the familiar
quasilinear flattening of the distribution function in the parallel
direction as a result of Landau damping.

In Figure~\ref{fig:ld_kaw}(b), we plot the corresponding
field-particle correlation $C_{E_\parallel} \left ( - (q_e
v_\parallel^2/2) \partial f_e(\V{v})/\partial
v_\parallel,E_\parallel\right)$ for the electrons using the same
correlation interval $\tau \omega_A = 10.0$.  One can see a similar
localization in velocity space of the energy transfer near the
resonant electron velocity $v_\parallel/v_{te}= \omega/(k_\parallel
v_{te})= 0.029$ (vertical black line), shown in more detail in
Figure~\ref{fig:ld_kaw}(c) where we have zoomed into the $v_\parallel$
range containing the resonant energy transfer. Also apparent in
Figure~\ref{fig:ld_kaw}(b) are two broader regions of energy transfer
at $0.5 \le |v_\parallel/v_{te}| \le 2.0$.  This component of the
energy transfer is odd in $v_\parallel$, and therefore cancels upon
integration over $v_\parallel$, leading to no net transfer of energy
between fields and particles. This component arises from the
incomplete cancellation of the larger-amplitude oscillating energy
transfer, both because the correlation interval $\tau$ is not exactly
an integral multiple of the wave period and because the damping of the
wave amplitude leads to incomplete cancellation in the second-half of
a wave period.  Note that performing the field-particle correlation
analysis at other spatial points in the simulation gives
qualitatively the same result.

A key point to emphasize about the field-particle correlation
technique is that the distribution of the energy transfer in velocity
space is expected to depend on the kinetic mechanism of energy
transfer. Other physical mechanisms---such as transit-time damping,
\cite{Barnes:1966} ion cyclotron damping,
\citep{Isenberg:2001,Isenberg:2012} stochastic ion heating,
\cite{Johnson:2001,Chen:2001,White:2002,Voitenko:2004,Chandran:2010a,Chandran:2010b,Klein:2016b}
or collisionless magnetic reconnection
\citep{Drake:2003,Pritchett:2004,Drake:2006,Egedal:2008,Egedal:2009,Parashar:2009,Egedal:2010,Markovskii:2011,Egedal:2012,Servidio:2012,Loureiro:2013,Dahlin:2014,Numata:2015}---are
expected to yield velocity-space signatures that are qualitatively
distinct from that of Landau damping.  Ongoing work shows that this
field-particle correlation technique, when an appropriate correlation
interval is chosen, still works in the presence of strong, broadband
kinetic plasma turbulence.\citep{Klein:2017b} In addition, the same
technique can be used to explore the transfer of free energy in
kinetic instabilities from unstable particle velocity distributions to
electromagnetic fluctuations. \citep{Klein:2017a}

\section{The Road Ahead}
The increasing availability of high cadence and high phase-space
resolution measurements of particle velocity distributions by
spacecraft and of gyrokinetic 3D-2V or fully kinetic 3D-3V nonlinear
simulations of weakly collisional heliospheric plasmas motivates a
concerted effort to develop new methods to maximize the scientific
return from these high-dimensional datasets. Plasma turbulence,
magnetic reconnection, particle acceleration, and instabilities are
four fundamental kinetic plasma processes operating under weakly
collisional conditions that significantly impact the evolution of the
heliosphere. The application of kinetic plasma physics to the study of
these heliospheric processes is primary driver of the new frontier of
\emph{kinetic heliophysics}.

Unlike in a traditional (strongly collisional) fluid, the removal of
energy from turbulent fluctuations and conversion of that energy to
plasma heat is a two-step process under the weakly collisional plasma
conditions relevant to many heliospheric environments. Learning to
handle the high-dimensional phase-space of kinetic plasma theory and
to exploit the information contained in velocity space holds the
potential for transformational progress in our understanding of
kinetic heliophysical processes.  The development of innovative
methods based on kinetic plasma physics, such as the field-particle
correlation technique highlighted here, will enable us to gain much
deeper insight into the dynamics and energetics of the heliosphere,
our home in the universe. And, beyond laying the foundation of
fundamental knowledge needed to construct a predictive capability for
heliophysics phenomena, such as extreme space weather, advances in our
understanding of fundamental physics through \emph{in situ}
measurements of heliospheric plasmas may be applied to better
comprehend the dynamics of more remote or extreme astrophysical
systems that lie out of reach of direct measurements.

Can we go further than some of the new directions discussed here to
exploit the information contained in velocity space of kinetic theory?
New capabilities enable fundamental aspects of the kinetic physics of
space plasmas to the explored in the laboratory under controlled or
reproducible conditions. The development of improved experimental
diagnostics to measure the particle velocity distribution functions
will enable some of the novel kinetic plasma physics methods endorsed
here to be applied in a laboratory setting.  Can machine learning,
coupled with sufficient physics insight from kinetic plasma theory, be
used to discover patterns in the high-dimensional phase space of
kinetic plasma theory? And, of course, our urgent need to understand
these essentially microphysical processes---turbulence, reconnection,
particle acceleration, and instabilities---is motivated by their
effect on the macroscopic evolution of the heliosphere, in particular
their impact on Earth and society.  Using our refined knowledge of
these kinetic physical mechanisms, we may attempt to build
next-generation models that couple their impact to the global
evolution of the heliosphere, enabling us to treat near-Earth space,
and other heliospheric environments, as complex systems. Efforts to
tackle the multi-scale problem of heliophysics through a rigorous
connection between the kinetic physics at microscales and the
self-consistent evolution of the heliosphere at macroscales will
propel the field of kinetic heliophysics into the future.

\begin{acknowledgments}
  I thank Dr. Kristopher Klein for his significant collaborative efforts
  in the development of the field-particle correlation method.  This
  work was supported by NSF PHY-10033446, NSF CAREER AGS-1054061, DOE
  DE-SC0014599, and NASA NNX10AC91G.  This work used the Extreme
  Science and Engineering Discovery Environment (XSEDE) , which is
  supported by National Science Foundation grant number ACI-1053575,
  through NSF XSEDE Award PHY090084.
\end{acknowledgments}

%


%

\end{document}